\documentclass[epj]{svjour}
\usepackage{graphicx}

\usepackage{rotating}
\usepackage[T1]{fontenc}
\usepackage[latin1]{inputenc}

\def\lapprox{{\raise0.5ex\hbox{$<$}\hskip-0.80em\lower0.5ex\hbox{$\sim$}

}}

\def\gapprox{{\raise0.5ex\hbox{$>$}\hskip-0.80em\lower0.5ex\hbox{$\sim$}

}}

\usepackage{graphics}
\begin{document}

\title{On the $\Sigma N$ cusp in the $pp \to pK^+\Lambda$ reaction}  

\author{
S.~Abd El-Samad\inst{1}\and
E.~Borodina\inst{2,3}\and
K.-Th.~Brinkmann\inst{4}\and
H.~Clement\inst{5,6} \and
E.~Doroshkevich\inst{5,6}\and
R.~Dzhygadlo\inst{2,3}\and
K.~Ehrhardt\inst{5,6}\and
A.~Erhardt\inst{5,6}\and
W.~Eyrich\inst{7}\and
H.~Freiesleben\inst{8}\and
W.~Gast\inst{2,3}\and
A.~Gillitzer\inst{2,3}\and
D.~Grzonka\inst{2,3}\and
C.~Hanhart\inst{2,3,9}\and
F.~Hauenstein\inst{2,3,7}\and
P.~Klaja\inst{2,3,7}\and
K.~Kilian\inst{2,3}\and
M.~Krapp\inst{7}\and
J.~Ritman\inst{2,3}\and
E.~Roderburg\inst{2,3}\and
M.~R\"{o}der\inst{2,3}\and
M. Schulte-Wissermann\inst{7}\and
W.~Schroeder\inst{2,3,7}\thanks{present address: Forschungszentrum J\"ulich,
  D-52428 J\"ulich, Germany}\and
T.~Sefzick\inst{2,3}\and
G.J.~Wagner\inst{5,6}\and
P. Wintz\inst{2,3}\and
P.~W\"{u}stner\inst{2,3,10}
}
%
\mail{H. Clement \\email: clement@pit.physik.uni-tuebingen.de}
%

\institute{
Atomic Energy Authority NRC Cairo, Egypt \and
Institut f\"ur Kernphysik, Forschungszentrum J\"ulich, D-52428 J\"ulich,
Germany \and
J\"ulich Center for Hadron Physics, Forschungszentrum J\"ulich, D-52428
J\"ulich, Germany \and
Physikalisches Institut Justus-Liebig-Universit\"at, D-35392
Gie$\beta$en,Germany \and
Physikalisches Institut der Universit\"at T\"ubingen, Germany \and
Kepler Center for Astro and Particle Physics,
 University of T\"ubingen, Auf~der~Morgenstelle~14, D-72076 T\"ubingen,
 Germany \and
Friedrich-Alexander-Universit\"at Erlangen-N\"urnberg, D-91058 Erlangen,
Germany \and
Institut f\"ur Kern- und Teilchenphysik, Technische Universit\"at Dresden,
D-01062 Dresden, Germany \and
Institute for Advanced Simulation, Forschungszentrum J\"ulich, D-52428
J\"ulich, Germany\and
Zentralinstitut f\"ur Elektronik, Forschungszentrum J\"ulich, D-52428
J\"ulich, Germany 
\\
(COSY-TOF Collaboration)}
\date{\today}
%
%
\abstract{
Measurements of the $pp \to pK^+\Lambda$ reaction at $T_p$ =
2.28~GeV have been carried out at COSY-TOF. In addition to the $\Lambda p$ FSI 
and $N^*$ resonance excitation effects a pronounced narrow structure is
observed in the Dalitz plot and in its projection on the $p\Lambda$-invariant
mass. The strongly asymmetric structure appears at the $pp \to NK^+\Sigma$ 
threshold and is interpreted as $\Sigma N$ cusp effect. 
The observed width of about 20 MeV/$c^2$ is substantially broader than
anticipated from previous measurements 
as well as theoretical predictions.
Angular distributions of this cusp structure
are shown to be dissimilar to those in the residual $pK^+\Lambda$ channel, but
similar to those observed in the $pK^+\Sigma^0$ channel.  
}
\maketitle
\section{Introduction}
\label{intro}

The hyperon production in nucleon-nucleon collisions has attracted interest
in recent years for a  number of reasons. First, it offers a valuable
tool for the determination of the hyperon-nucleon final state interaction
(FSI). Most directly this 
effect is seen in the hyperon-nucleon invariant mass spectrum. For a recent
determination of the $\Lambda p$-FSI by the COSY-HIRES collaboration using a
high-resolution magnetic spectrometer see Ref.~\cite{hires1}. Second, it gives
 access to the rare decay branches of $N^*$ resonances produced
in the $NN$ collisions.
 For recent work on that see Refs.~\cite{DD,DD12,ER,ERalt1,ERalt2}. 
Third, it offers the chance to search for more exotic objects like pentaquarks 
\cite{TOF1,TOF2,ANKE} or dibaryons \cite{hires2,DISTO}. 
Among the latter an
inevitable, though not very exotic candidate would be the $\Sigma N$
system produced near threshold followed by a $\Sigma N \to \Lambda N$
transition. Whenever an inelastic channel opens in a production reaction, 
it produces a non--analyticity in the spectra, a so-called cusp, the
strength of which is a measure of the corresponding transition matrix 
element (here  $\Sigma N\to \Lambda N$) - {\it e.g.} see Ref.~\cite{gasser} for 
the method to extract the s-wave $\pi\pi$ scattering lengths from $K\to 3\pi$.
 Signs of such a cusp in the $pp \to pK^+\Lambda$ 
reaction have been first observed in inclusive measurements with single-arm
magnetic spectrometers at Saclay \cite{sieb} and COSY-HIRES
\cite{hires3}. A theoretical calculation of this cusp effect in
proton-proton collisions has been presented by Laget~\cite{Laget}.

Originally a peak structure at the position of the  $\Sigma N$ threshold has
been discovered in $K^-$ absorption in deuterium \cite{dahl}.  Subsequent 
bubble-chamber measurements  \cite{tan,braun} observed a pronounced peak at
the $\Sigma$ threshold with a width of about 10 MeV/$c^2$ and below. For a
review and a discussion of possible dibaryon aspects see, {\it e.g.}
Refs.~\cite{Dalitz,bad,gal,torres}. 

In exclusive and kinematically complete COSY-TOF measurements 
indications of this cusp have been observed at several energies
\cite{ER,ERalt1,ERalt2}. Since these measurements lack the necessary
statistics for a reliable investigation of the cusp effect, we use here the
measurements at $T_p$ = 2.28~GeV (p = 3.081~GeV/c) for a detailed
investigation of this matter. The primary purpose of this run, 
which comprises an order of magnitude higher statistics than the previous
TOF-measurements,
 was originally the pentaquark search
\cite{TOF1}. Angular distributions obtained from this run have been published
already for both $pK^+\Lambda$ and $pK^+\Sigma^0$ channels \cite{DD}. 
The results presented in this paper are based on the thesis work of
Refs. \cite{KE,MR}, where also details of experiment and analysis are found.
\section{Experiment}
\label{sec:2}

\subsection{Detector setup}

Since the experimental setup was discussed in detail already in
Refs.~\cite{DD,ER,TOF1}, we give here only a short account. 
The measurements were carried out at the J\"ulich Cooler Synchrotron COSY
using the time-of-flight spectrometer TOF located at one of its external beam
lines. 
The TOF spectrometer is a modular detector setup, which can be adapted to the
specific requirements of an experiment. Here it 
was used in its standard version for hyperon
production, see {\it e.g.}~Figs.~1 and 2 of Ref.~\cite{ER}. At
the entrance to the detector system the beam -- focused to a diameter smaller
than 2 mm -- hits the thin-walled LH$_2$ target, which has a length of 4 mm, a
diameter of 6 mm and 0.9 $\mu m$ thick hostaphan foils as entrance and exit
windows.  
At a distance of 22 mm downstream of the target the two layers of 
the start detector (each consisting of 1 mm thick scintillators cut into 12
wedge-shaped sectors) were placed.
 A silicon microstrip detector as well as 
two fiber hodoscopes were installed at distances 30, 100 and
200 mm from the target. These three tracking detectors provide the position
information of the traversing charged particles,  
whereas the  start detector supplies the start time signals for the
time-of-flight (TOF) measurements.
After a flight path of about 3~m through the evacuated vessel the charged
particles are detected in the highly segmented stop detector system consisting
of the triple-layered quirl and ring detectors as well as a single-layered
96-fold segmented barrel detector.  

\subsection{Particle identification and event reconstruction}

In the experiment the trigger suitable for the selection of hyperon production
events required two hits in the start detector and four hits in the
stop detectors. This multiplicity jump from two to four specifically selects
the production of neutral hyperons, which decay into charged products like the
$\Lambda$ decay process $\Lambda \to p\pi^-$, which happens with a branching
fraction of 64$\%$.  
Tracks of charged particles are reconstructed from straight-line fits to the
hit detector elements in start, fiber and stop detectors. They are accepted as
good tracks of primary particles ($p, K^+$), if they originate in the target.
The secondary particles resulting from hyperon decay ($p, \pi^-$) form a
$V$-shaped track pair originating from a secondary vertex downstream of the 
microstrip detector. 
Primary vertices, which are located within the target volume, were
reconstructed with an accuracy of $\sigma_{x,y}$ = 0.25 mm and $\sigma_z$ = 0.7
mm. The secondary vertex from the $\Lambda$ decaying downstream the microstrip
detector and upstream the first hodoscope was reconstructed with accuracy of
$\sigma_{x,y}$ = 1.5 mm and $\sigma_z$ = 4.5 mm.

In addition to the tracking information we use the TOF information of all four
ejectiles of an event to determine their four-momentum vectors. For the
particle identification a kinematic fit is applied, where all permutations of
particle assignments are considered and the one with the best $\chi^2$ is
selected as the correct one. 
Since the kinematics of the light($\pi, K$) and heavy ($p$) emitted particles 
is quite different, finding of the correct particle assignments actually is
rather clear-cut. According to detailed  Monte Carlo (MC) simulations the
misidentification rate was in the order of a few percent only -- in agreement
with previous TOF results \cite{DD,ER}.  

\begin{figure}
\begin{center}
\includegraphics[width=20pc]{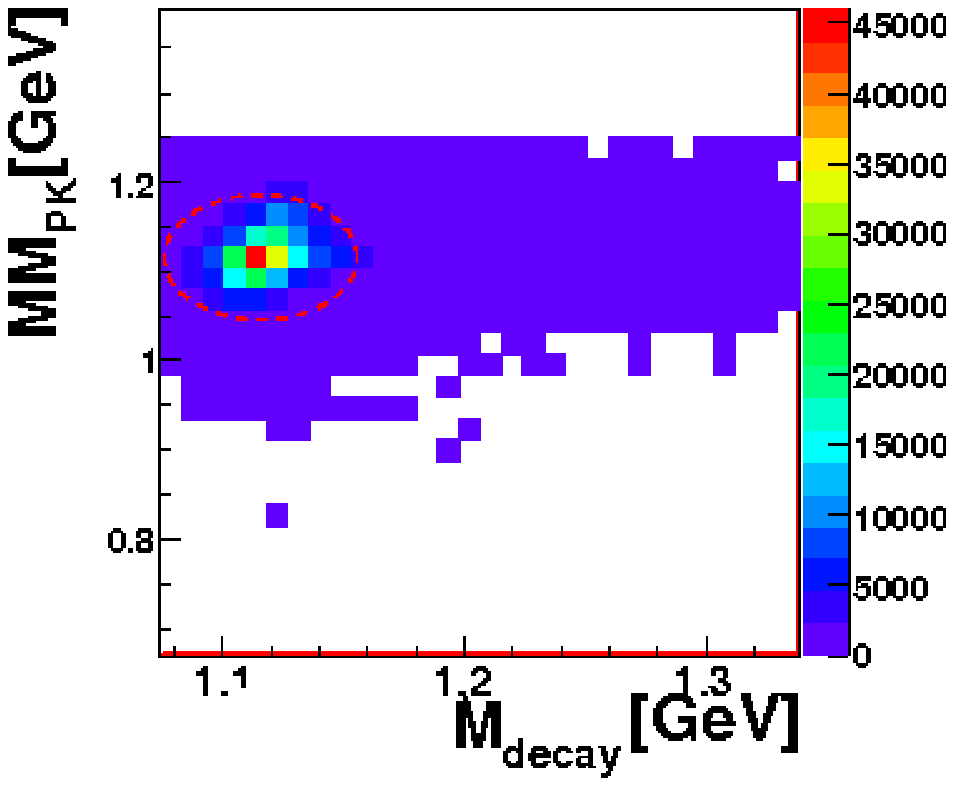} 
\includegraphics[width=20pc]{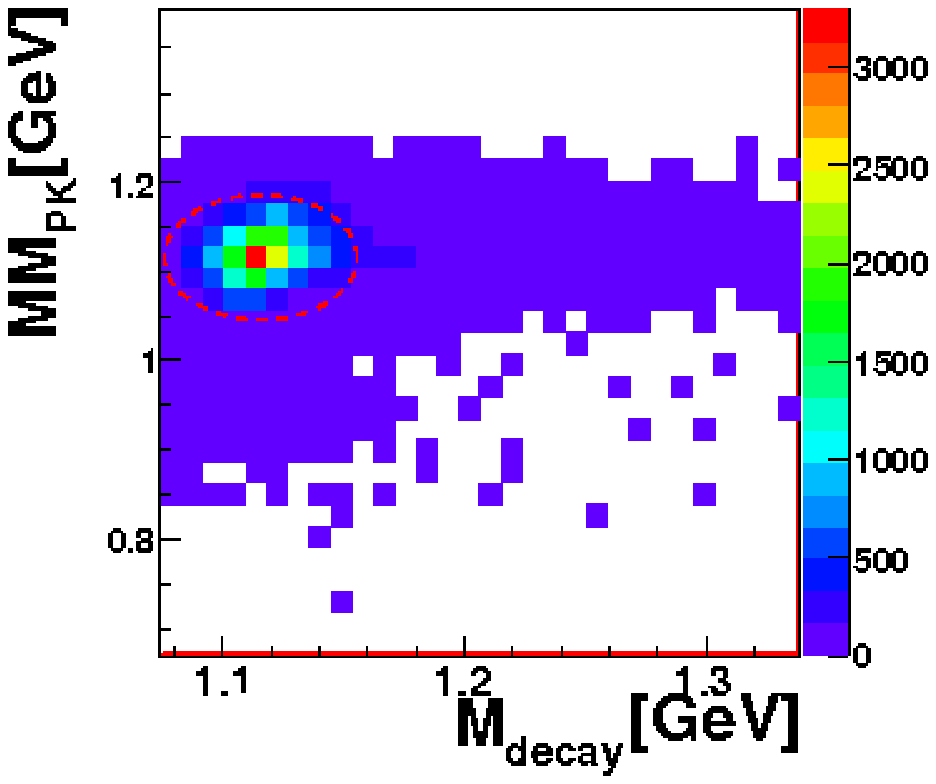} 
\end{center}
\caption{ Plot of the invariant mass $M_{p\pi^-} := M_{decay}$ of the $\Lambda$ decay
  particles versus the primary particle $p K^+$-missing mass $MM_{pK}$ for MC
  simulation ({\bf top}) and data ({\bf bottom}). The dashed circles indicate
  the range accepted for the subsequent kinematic fit. The scale of the z-axis
  is in arbitray units.
}
\end{figure}

Fig.~1 depicts the two-dimensional plot of the invariant mass $M_{p\pi^-}$ of
the $\Lambda$ decay particles versus the primary particle $p K^+$-missing mass
spectrum for MC simulation (top) and data (bottom). As can be seen the
resolution in invariant and missing mass are comparable, and the MC
simulation reproduces very well the experimental situation. The dashed circles
in Fig.~1 indicate the region of events accepted for the subsequent analysis
steps.  

The kinematic fit was fivefold
overconstrained, where the fifth overconstraint originates from the
condition that the invariant mass of the four-momenta of the decay particles
has to be identical to the $\Lambda$ mass.
A total of 30,000 events passed the criterion of $prob(\chi^2) > 10 \%$ in the
kinematic fit for being accepted as a proper event \cite{KE}. 

The primary particle $p K^+$ missing mass spectrum of the events finally
selected by the $\chi^2$ criterion is shown in Fig.~2 {\it before} the
kinematic fit. The $\Lambda$ peak appears with essentially no background and
is in very good agreement with the Monte Carlo (MC) simulations of the
detector performance. The slight mass shift of the peak between data and MC is
due to imperfections in the energy calibration.

The MC simulations also have been used for efficiency and
acceptance corrections of the data. We used
for these simulations as input a model-description for the reaction of
interest, which provides a good description of the experimental differential
distributions -- see discussion in section 4. We note, however, that since the
TOF detector covers nearly the full phase space of the reaction -- see the
Dalitz plots in Fig.~3 -- the differences in the corrections between
model-based and pure phase MC simulations are only minor.

By use of the kinematic fit the mass resolution in the spectra of the
invariant masses $M_{K\Lambda}$, $M_{pK}$ and $M_{p\Lambda}$ (Fig.~4) improves from
about 30 MeV/$c^2$ to 6 MeV/$c^2$ (FWHM). 

\begin{figure}
\begin{center}
\includegraphics[width=17pc]{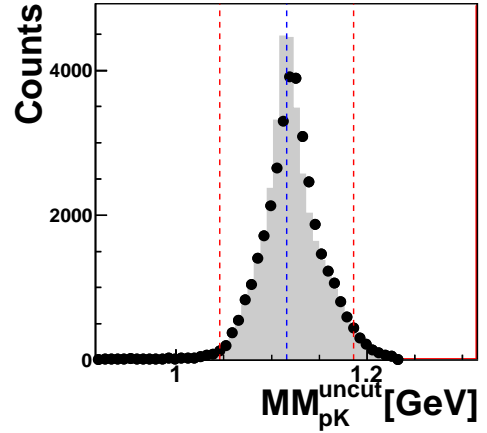} 
\end{center}
\caption{ $p K^+$-missing mass spectrum of events, which were finally selected
  as good candidates for the $pK^+\Lambda$ channel by the $\chi^2$
  criterion. The shaded area gives the MC simulation, the vertical middle line
  the position of the $\Lambda$ mass and the two outer vertical lines the
  range accepted for the subsequent kinematic fit.
}
\end{figure}

The absolute cross section is obtained by relative normalization to literature
data \cite{edda,said} for the elastic $pp$ channel, which was measured in
parallel during the experiment.

\section{Results}
\label{sec:3}

For the total cross section of the $pp \to pK^+\Lambda$ reaction at 2.28 GeV
we obtain  $(21.2\pm 0.2\pm 2.0)$ $\mu$b,
where the errors are statistical and systematic respectively. 
Within uncertainties this value agrees with that obtained in
Ref.~\cite{DD}. 
As in the previous TOF results \cite{DD,ER,ERalt1,ERalt2} for the $pp \to
pK^+\Lambda$ reaction the systematic uncertainty of about 10$\%$  is the by
far dominant uncertainty. It originates from the uncertainties in the
luminosity determination, reconstruction efficiency and acceptance correction. 

Single differential cross sections
are shown in
Figs. 4 - 6. Fig. 3 exhibits the acceptance and efficiency corrected 
Dalitz plot of the three-body exit channel $pK^+\Lambda$. It is by no means
homogeneous, {\it i.e.}~phase space-like.
The intensity in the Dalitz plot peaks at the left side of the short diagonal
corresponding to high $p K^+$-invariant masses $M_{pK^+}$. In 
addition we see two vertical narrow structures corresponding to
the $p\Lambda$-invariant masses $M_{p\Lambda}$ at threshold and at
$M_{p\Lambda}^2 \approx$ 4.54~GeV$^2/c^4$ (see vertical arrows in Fig.~3), {\it
  i.e.} $M_{p\Lambda} \approx$ 2.13~GeV$/c^2$ $\approx m_p + m_{\Sigma}$. The
first structure may be  related to the $\Lambda p$ FSI, whereas the latter one
is in the region of the $\Sigma p$ production threshold.

\begin{figure} [t]
\begin{center}
\includegraphics[width=20pc]{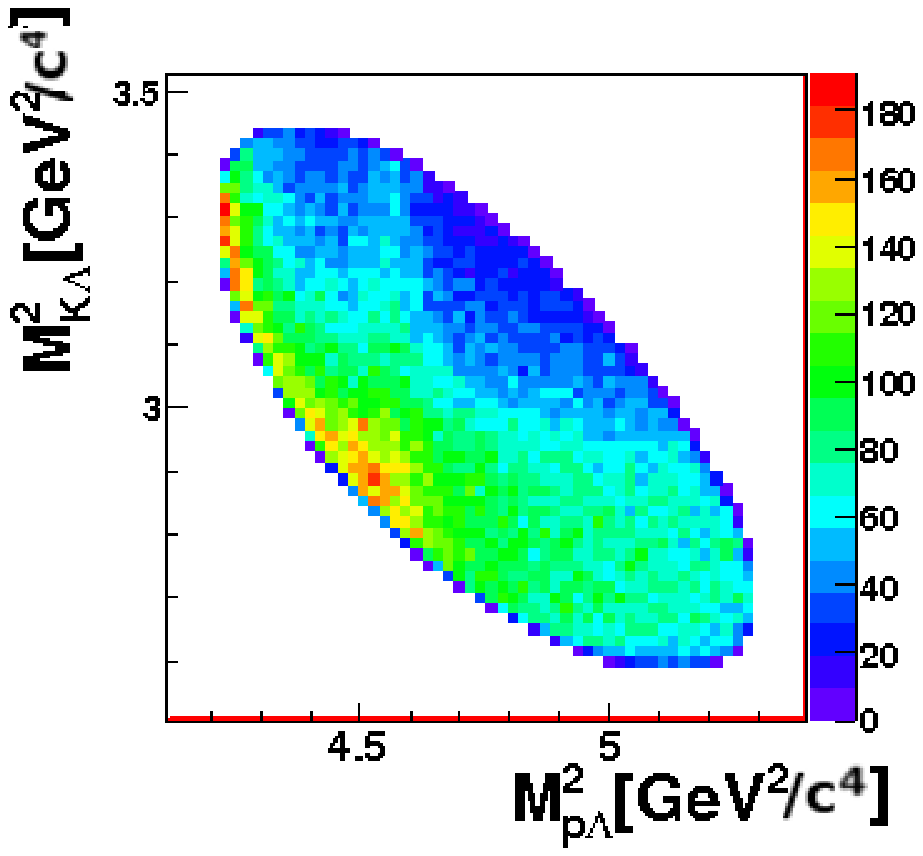} 
\includegraphics[width=20pc]{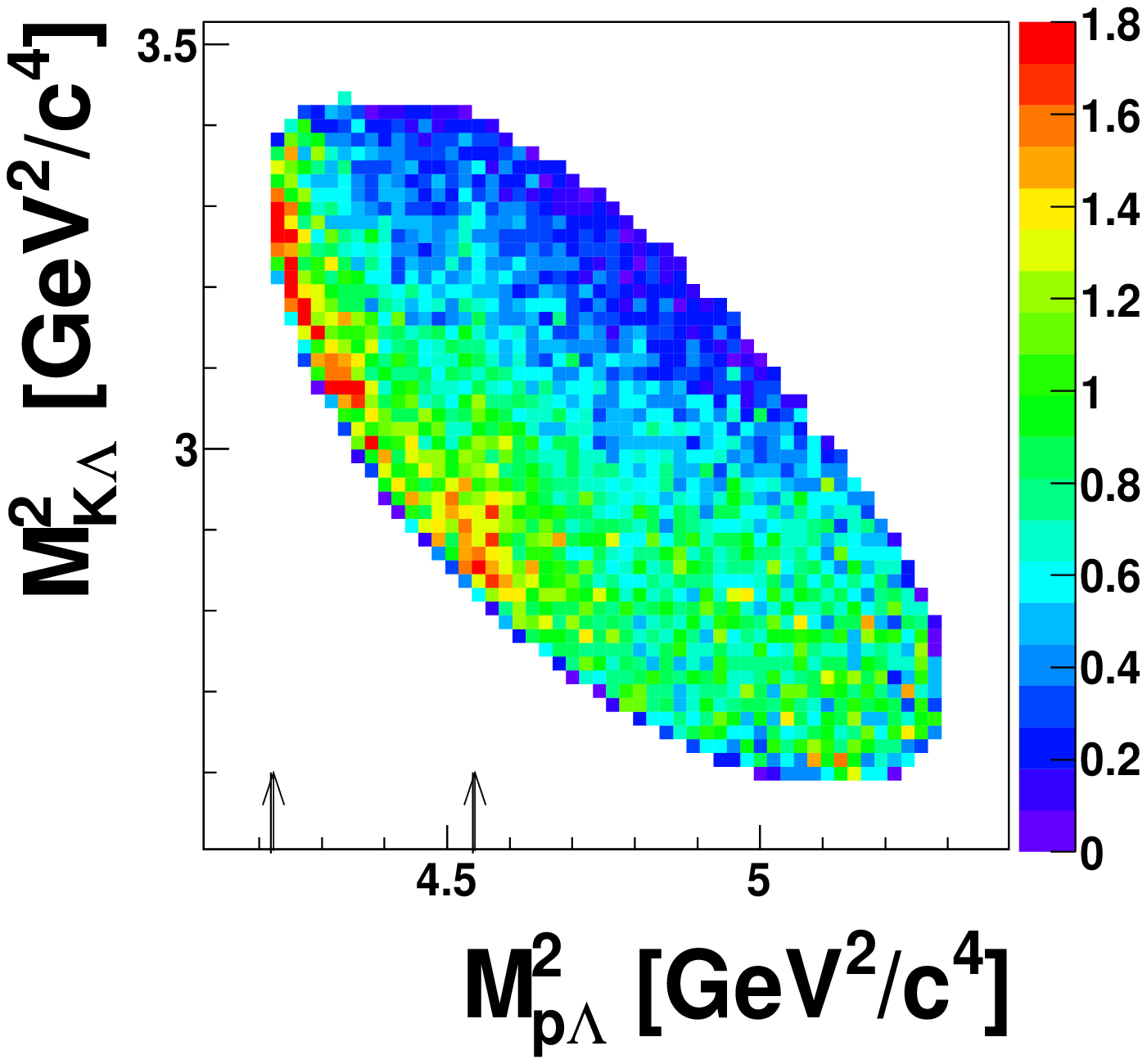}
\end{center}
\caption{Dalitz plot of $M_{K^+\Lambda}^2$ versus  $M_{p\Lambda}^2$ for the
  $pp \to pK^+\Lambda$ reaction at $T_p$ = 2.28~GeV, for MC simulation ({\bf
  top}) and data ({\bf bottom}). 
  The two vertical narrow structures (see arrows)  at the $p\Lambda$ threshold
  and at $M_{p\Lambda}^2$ = 4.54~GeV$^2/c^4$ are due to the $p\Lambda$ FSI and
  the $\Sigma N$ cusp, respectively. The scale of the z-axis is in arbitray
  units. 
}
\end{figure}


\begin{figure} 
\begin{center}
\includegraphics[width=18pc]{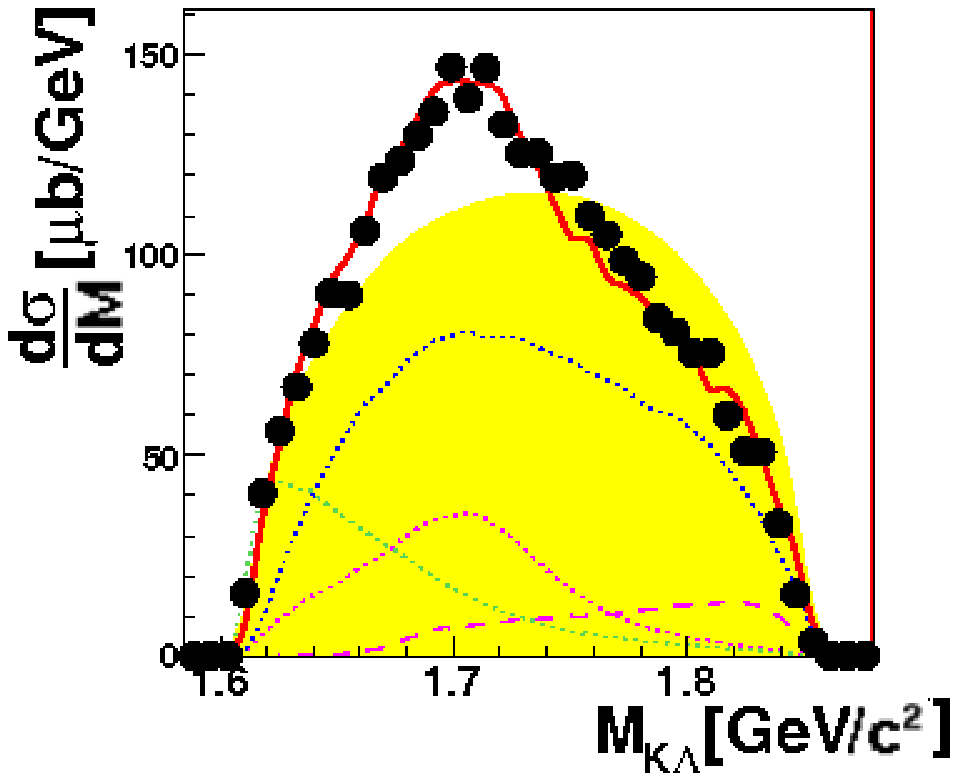}
\includegraphics[width=18pc]{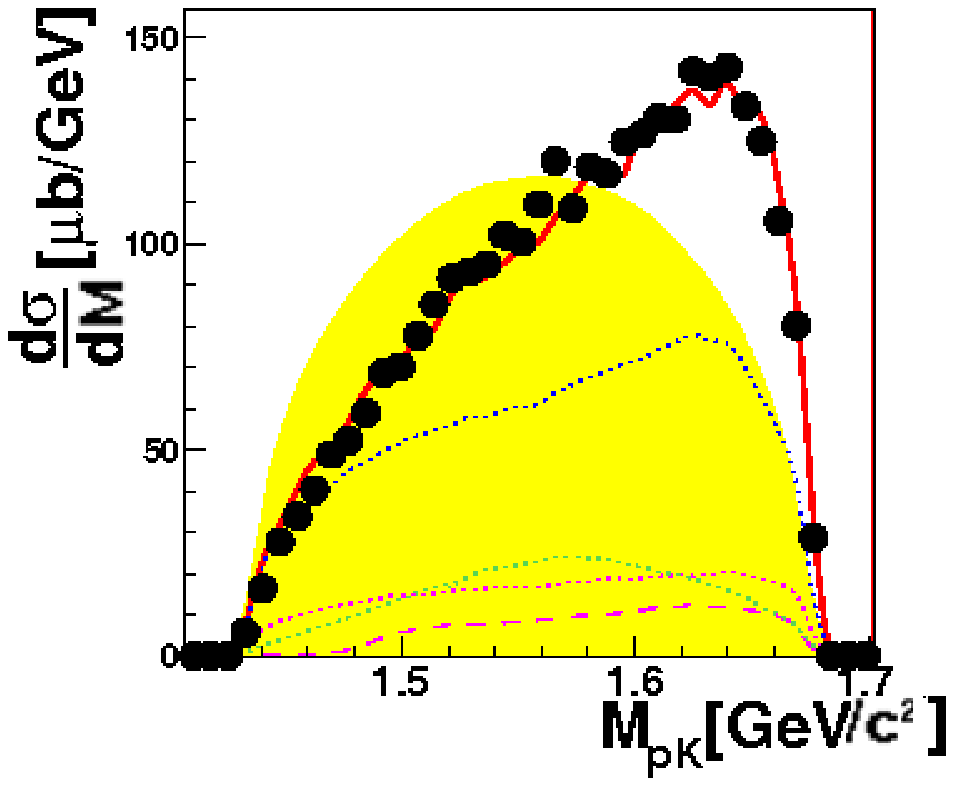}
\includegraphics[width=18pc]{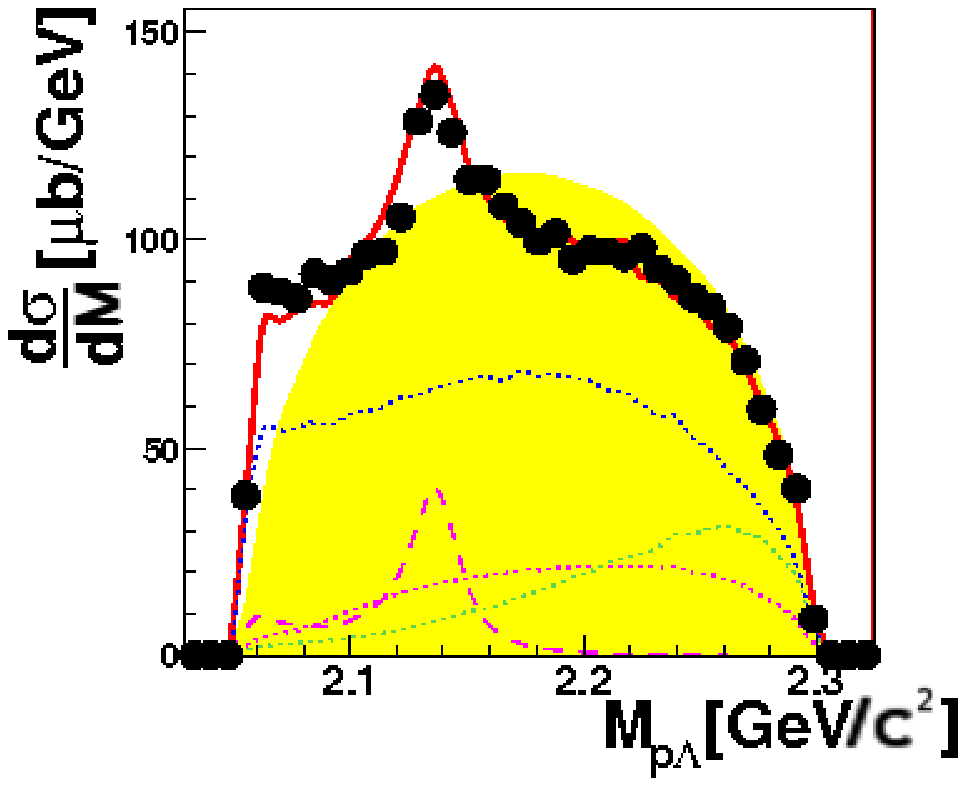} 
\end{center}
\caption{Differential distributions of the invariant-mass systems
  $M_{K^+\Lambda}$ ({\bf top}), $M_{pK^+}$ ({\bf middle}) and  $M_{p\Lambda}$ ({\bf
    bottom}). The shaded areas indicate phase-space distributions, the dotted
  lines the contributions of $N^*$ resonances, the dashed line the $\Sigma$
  cusp effect and the solid line the full MC simulation.
}
\end{figure}


In Fig. 4 we show the spectra of the three invariant mass systems
$M_{K\Lambda}$, $M_{pK}$ and $M_{p\Lambda}$
The data of all three strongly deviate from phase space, which is indicated by
the shaded area in the plots. Among them the $M_{p\Lambda}$ spectrum appears
particularly interesting, since it exhibits two narrow structures. They may be 
connected with FSI and $\Sigma$N cusp as mentioned above 
and will be discussed in the following.

\section{Discussion}
\label{sec:4}

As shown in previous works \cite{DD,ER,ERalt1,ERalt2}, the gross features of
the  $pp \to pK^+\Lambda$ reaction  may be well described by $N^*$
resonance production with subsequent $N^* \to K^+\Lambda$ decay and by 
inclusion of the $\Lambda p$ FSI. 
The $N^*$ resonances, which play a role here, are $N(1650)1/2^-$,
$N(1710)1/2^+$ and $N(1720)3/2^+$.
For the description of the data we assume excitation of these $N^*$
resonances via $t$-channel meson exchange along the prescription given in
Refs. \cite{ER,KE}. The $\Sigma$ cusp is treated in a simplified manner 
just as a narrow $\Lambda p$ resonance and the $\Lambda p$ FSI is taken into
account in the factorization approximation of Ref. \cite{hires1}. 

For the model fit to the data we allow mass and width of the $N^*$
resonances to vary within the boundaries given in PDG \cite{pdg}. In addition
relative phase and strength parameter for each of the resonances have been
fitted. For a quantitative description of the
data in the FSI region we also need a readjustment of the FSI parameters
resulting in $a_{\Lambda p}$ = -2.2 fm and $r_{\Lambda p}$ = 1.4 fm. These
values are within $1 \sigma$ and $2 \sigma$, respectively, of the HIRES result
\cite{hires1}.  
\begin{table}
\caption{Results from the fit to the data for $N^*$ resonances.
}   
\begin{tabular}{lllllll} 
\hline


 &resonance& $m_{N^*}$ & $\Gamma_{N^*}$ &$\Phi_{N^*}$ &$\sigma$ \\ 
 &            & (MeV) & (MeV) & (deg) & $\mu$b \\
\hline

& $N(1650)1/2^-$ & 1653 & 168 & 29 &  ~3.5(4) \\
& $N(1710)1/2^+$ & 1712 & ~99 & 16 &  ~3.5(4) \\
& $N(1720)3/2^+$ & 1731 & 383 & ~0 &  12.8(13) \\
\hline
 \end{tabular}\\
\end{table}

The result of a fit to the data is shown in Fig. 3, top, as Dalitz plot
and in Fig.~4 by dotted ($N^*$ resonances), dashed ($\Sigma$ cusp) and
solid (full calculation) lines. The resulting values for mass, widths and 
relative phase of the $N^*$ resonances and $\Sigma$ cusp are given in Table 1
in addition to their total cross section contributions. The data are
reasonably well described by this fit. According to this analysis the
enhancement (relative to phase space) in the $M_{p\Lambda}$ spectrum at the
$p\Lambda$ threshold is predominantly due to the $\Lambda p$ FSI, whereas the
enhancement at high $p\Lambda$ masses arises from $N^*$ excitation, in
particular from the excitation of $N(1650)1/2^-$ -- see Fig.~4.
A much more sophisticated Dalitz plot analysis in the
framework of Ref.~\cite{sib} taking into account all COSY-TOF data on this
reaction is in progress. Hence we will not discuss the contribution of $N^*$
resonances here further, but rather concentrate on the discussion of the cusp
effect.  The primary goal of the fit here is just to have a reasonable
description of the data for the purpose of a reliable acceptance and
efficiency correction by MC simulations -- as discussed above in section 2 --
and to also have some reliable estimate of the physical background (due to
$N^*$ resonances) underneath the $\Sigma$ cusp. Also, as we see from the
fit in Fig.~4, bottom, the assumption of a Breit-Wigner distribution for the
$\Sigma$ cusp is not a good description of this phenomenon.

\begin{figure} 
\begin{center}
\includegraphics[width=20pc]{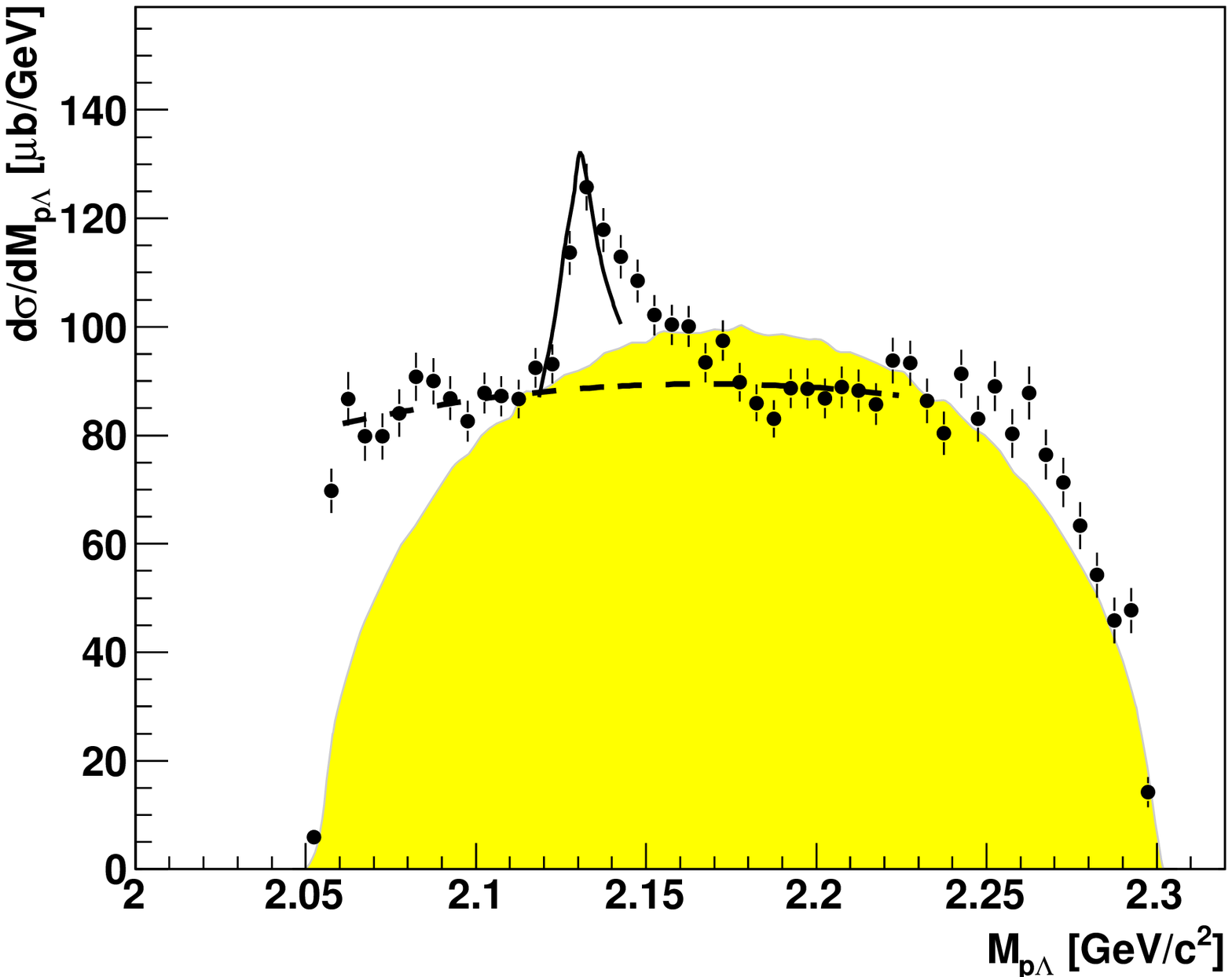}
\includegraphics[width=20pc]{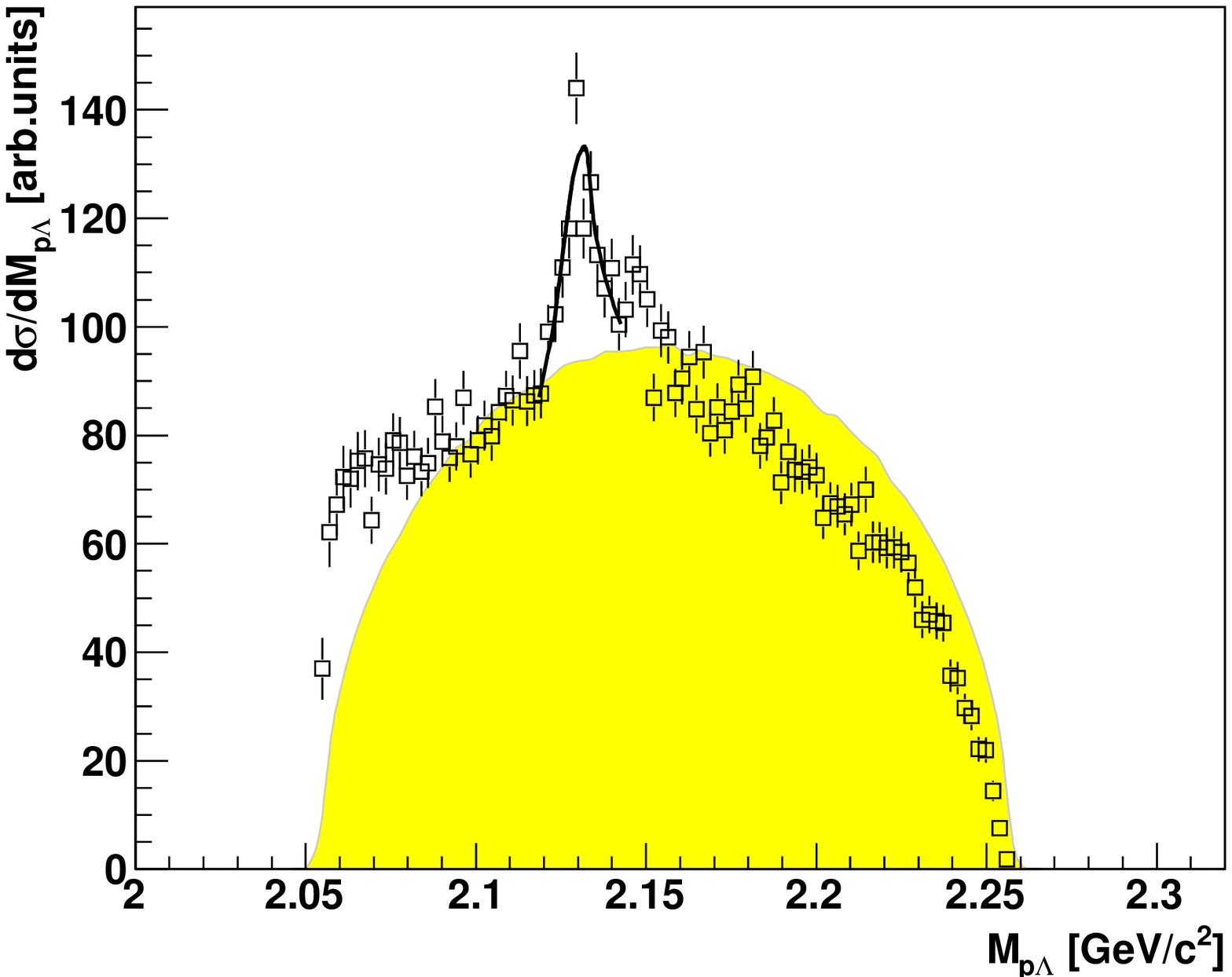}
\includegraphics[width=20pc]{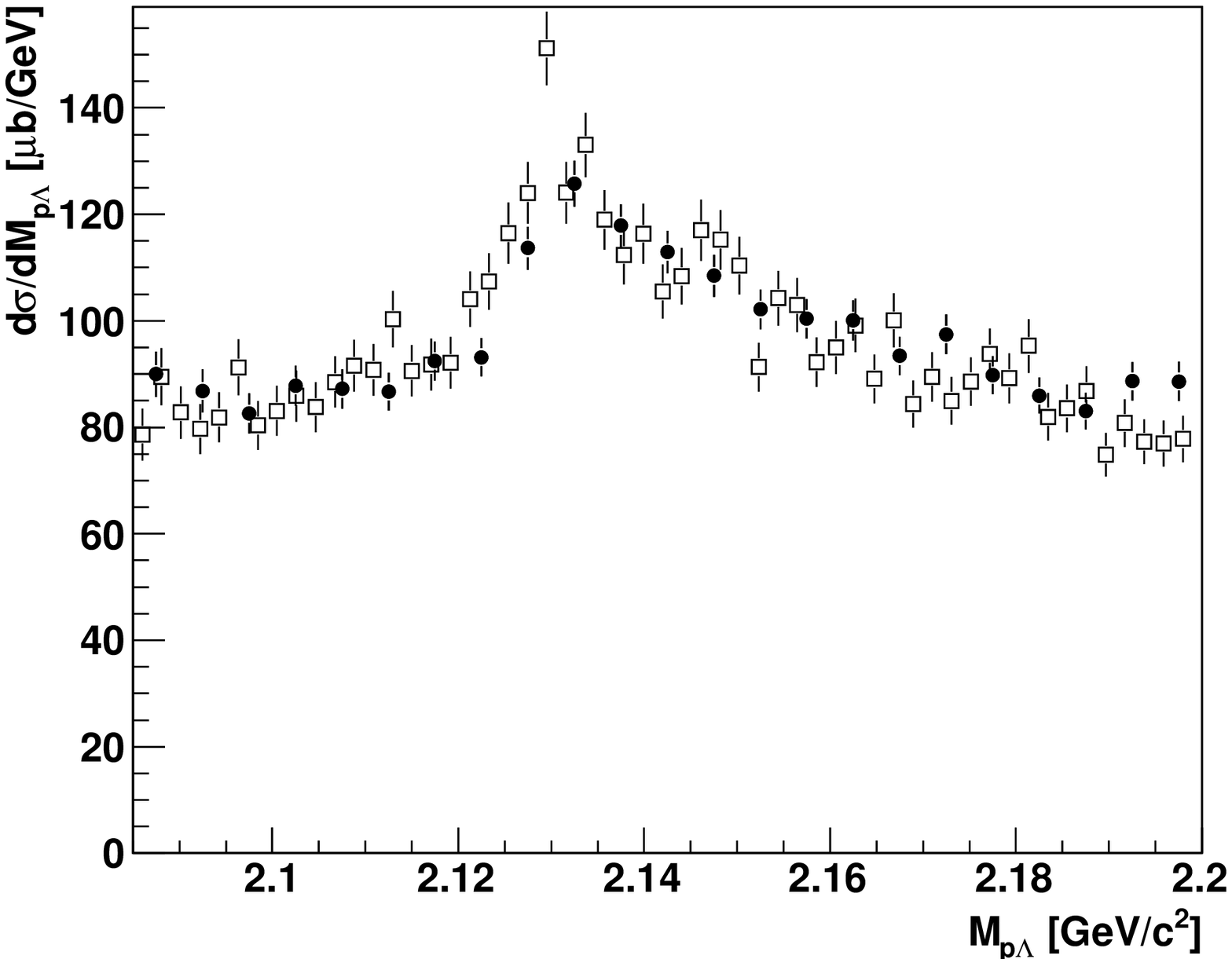}
\end{center}
\caption{Differential distribution of the invariant-mass system
  $M_{p\Lambda}$. The shaded area represents the phase-space distribution, the
  solid  curve shows the shape of the cusp as obtained from a coupled-channel
  treatment \cite{sascha}, averaged over the experimental resolution and fitted
  in height to the data at the low-energy side of the cusp. The cusp is
  assumed to sit upon a smooth background represented by the dashed line.  
  {\bf Top}: data of this work taken at 3.081 GeV/c. {\bf Middle}: data
  taken with the new straw tracker at 2.95 GeV/c, normalized arbitrarily
  \cite{MR}. {\bf Bottom}:  comparison of both data sets in the cusp region.   
}
\end{figure}

In fact, since the cusp is a threshold phenomenon we do not expect a symmetric
Breit-Wigner distribution, but an asymmetric energy
dependence in form of a Flatt\'{e} distribution \cite{Flatte} 
%
\begin{center}
$d\sigma / d M_{\Lambda p}~\sim~ \Gamma_{\Lambda p}~/~|m_R^2 -
  m_{\Lambda p}^2 - i m_R (\Gamma_{\Lambda p} + \Gamma_{\Sigma p})|^2~$
\end{center}
with
\begin{center}
$\Gamma_{\Lambda p} = g_{\Lambda p} q_{\Lambda p}$ and 
$ \Gamma_{\Sigma p} = g_{\Sigma p} q_{\Sigma p},$
\end{center}
where $g_i$ and $q_i$ are coupling constants and cm momenta, respectively, in
the corresponding two-body subsystems. We have 
\begin{center}
$q_{\Sigma p} = \frac{ \sqrt{(m_{\Sigma p}^2 -(m_{\Sigma} +m_p)^2)
               (m_{\Sigma p}^2 - (m_p - m_\Sigma)^2)}}{2m_{\Sigma p}}$ 
\end{center}
and
\begin{center} 
$q_{\Sigma p} =  i \frac{ \sqrt{((m_{\Sigma} +m_p)^2 -m_{\Sigma p}^2)
               (m_{\Sigma p}^2 - (m_p - m_\Sigma)^2)}}{2m_{\Sigma p}}$ 
\end{center}
above and below threshold, respectively.

%
If $g_{\Sigma N} \ll g_{\Lambda p}$ then the Flatt\'{e} distribution approaches
a symmetric distribution. If $g_{\Sigma N} \gg g_{\Lambda p}$   we
have a very asymmetric distribution with a trailing slope at energies below
the cusp and a rapid decline beyond the cusp. This situation is just
opposite to what 
we observe in our data. Also, a large $g_{\Sigma N}$ means a strong $\Sigma N$
FSI, which in turn causes a strong low-mass enhancement in the $\Sigma N$
invariant-mass spectrum as well as a steep increase of the total cross section
near threshold of the $pp \to \Sigma N K$ reactions. Explicit measurements of 
these $\Sigma$ production channels \cite{DD,DD12} provide no evidence for
that. Since we have a sizeable $\Lambda p$ FSI, we hence 
expect $g_{\Sigma N} \ll g_{\Lambda p}$, {\it i.e.} a more or less symmetric
distribution around the cusp. This is borne out also in a recent theoretical
treatment of $\Lambda$ and $\Sigma$ production in $NN$ collisions in the
framework of the coupled channel effective range method \cite{sascha}. The
resulting cusp distribution folded with the appropriate experimental
resolution is shown in Fig. 5 as solid curve. Similar to the prediction of
Laget~\cite{Laget} we have the situation that the low-energy side agrees very
well with the data, whereas the data on the high-energy side fall off much
less steep than predicted.

%
%

In order to check whether the observed larger width is due to our
experimental resolution, we compare in Fig. 5 our $M_{p\Lambda}$ data from the
high-statistics run with a more recent COSY-TOF high-resolution measurement at
2.95 GeV/c \cite{MR}, see Fig.~5, middle. This measurement was performed
utilizing the new  straw-tracker at COSY-TOF providing an invariant mass
resolution of 2.6 MeV/$c^2$ FWHM, {\it i.e.} more than twice better than in
this work. Both data sets are compared in Fig.~5, bottom, depicting the cusp
region in enlargement. Aside from a possible slight shift of about 1 - 2
MeV/c$^2$, which is within the uncertainty of mass calibrations, both data sets
coincide -- in particular, if the different resolutions are taken into account.
We see that both data sets exhibit the cusp in compatible shape and
also size (relative to the background). 

Admittedly, there appears to be possibly a slight difference. Whereas the
high-statistics data exhibit a gentle decline at the high-energy side, the
high-resolution data show the indication of a roughly 5 MeV broad bump upon
the declining slope near 2.146 GeV/$c^2$. Since this bump effect may be less
than $3\sigma$ -- depending on the assumption of background -- and hence not
statistically significant, we do not want to speculate about its nature at
this point. However, we would like to mention that already kaonic deuterium
data suggested a two-bump scenario. In particular the bubble-chamber data with
the highest statistics and a quoted mass resolution of $\sigma$ = 1.0 - 2.6
MeV/$c^2$ \cite{tan} have been fitted by two symmetric Breit-Wigner functions
with $m_1 = 2128.7\pm0.2$  MeV/$c^2$, $\Gamma_1 = 7.0\pm0.6$  MeV/$c^2$ and
$m_2 = 2138.8\pm0.2$  MeV/$c^2$, $\Gamma_2 = 9.1\pm2.4$  MeV/$c^2$ suggesting
two nearby resonance states separated by 10 MeV/$c^2$. The position of this
second bump (shoulder) is, however, not compatible with the position of the
unincisive bump at 2.146 GeV/$c^2$ in Fig. 5, middle and bottom. From this we
conclude that at present there is no statistically solid evidence
for a second bump beyond the $\Sigma N$ cusp, however, there is a solid
evidence for a surplus of cross section right beyond the cusp position, which
so far is not understood theoretically.  

The situation might change and the physical relevance of the second bump
structure seen in Fig.~5, middle, might have to be re-discussed, if it should
turn out that the cusp effect has indeed an energy dependence at the
high-energy side as given by Refs. \cite{Laget,sascha}. Having a width of only
a few MeV, such a bump appears to be smeared out in our measurement with
the coarser energy resolution shown in Fig.~5, top, producing there just a
shoulder. Also, the differences between the two-resonance scenario of Tan {\it
  et al.} \cite{tan} and our results for cusp and second bump could be
reconciled, if we allow to shift their resonance masses by 3 MeV/$c^2$, so
that their first resonance coincides with the value for the $\Sigma N$
threshold -- as we find it for the position of the cusp in our data.  That way
the value for the second resonance mass in Ref. \cite{tan} would move to 2.142
GeV/$c^2$ and be no longer in serious disagreement with the position of the
second bump in our high-resolution spectrum. For a new reinvestigation of this
two-resonance scenario including the data of this work we refer to
Ref. \cite{machner}. 

The different behavior at the high-mass end  of the two data sets
displayed in Fig.~5 is just due to the different beam energies: For
the 2.95~GeV/c data the kinematic phase-space limit is 2.258~GeV/$c^2$, for
the 3.08~GeV/c data the corresponding value is 2.299~GeV/$c^2$. In addition the
influence of the broad $N^*$ resonances, which determine the continuum below
the cusp, depends on the beam energy -- as demonstrated in Ref. \cite{ER}.
%

As expected the cusp is located at $m_p + m_{\Sigma^0}$ = 2131 MeV/$c^2$.
This is consistent with the assumption that the observed structure corresponds 
to the production of $\Sigma$ right at its threshold. In
principle there should be two $\Sigma N$ cusps, since $m_p + m_{\Sigma^0} =
m_n + m_{\Sigma^+}$ + 2 MeV/$c^2$. However, due to the finite invariant-mass
resolution of the data we are unable to separate these. 
Since the excitation of $N^*$ resonances gives a flat energy dependence in the
$M_{p\Lambda}$ distribution, see Fig.~4,
we assume the physical background not belonging to
the cusp scenario to be represented by the dashed curve drawn in
Fig.~5. Associating that way the surplus of cross section above the dashed
line with the cusp scenario, {\it i.e.} neglecting possible interference
effects, we obtain a  
cusp cross section, which corresponds to roughly 5$\%$ of the total $pp
\to pK^+\Lambda$ cross section. 

\begin{figure}[t]
\begin{center}
\includegraphics[width=17pc]{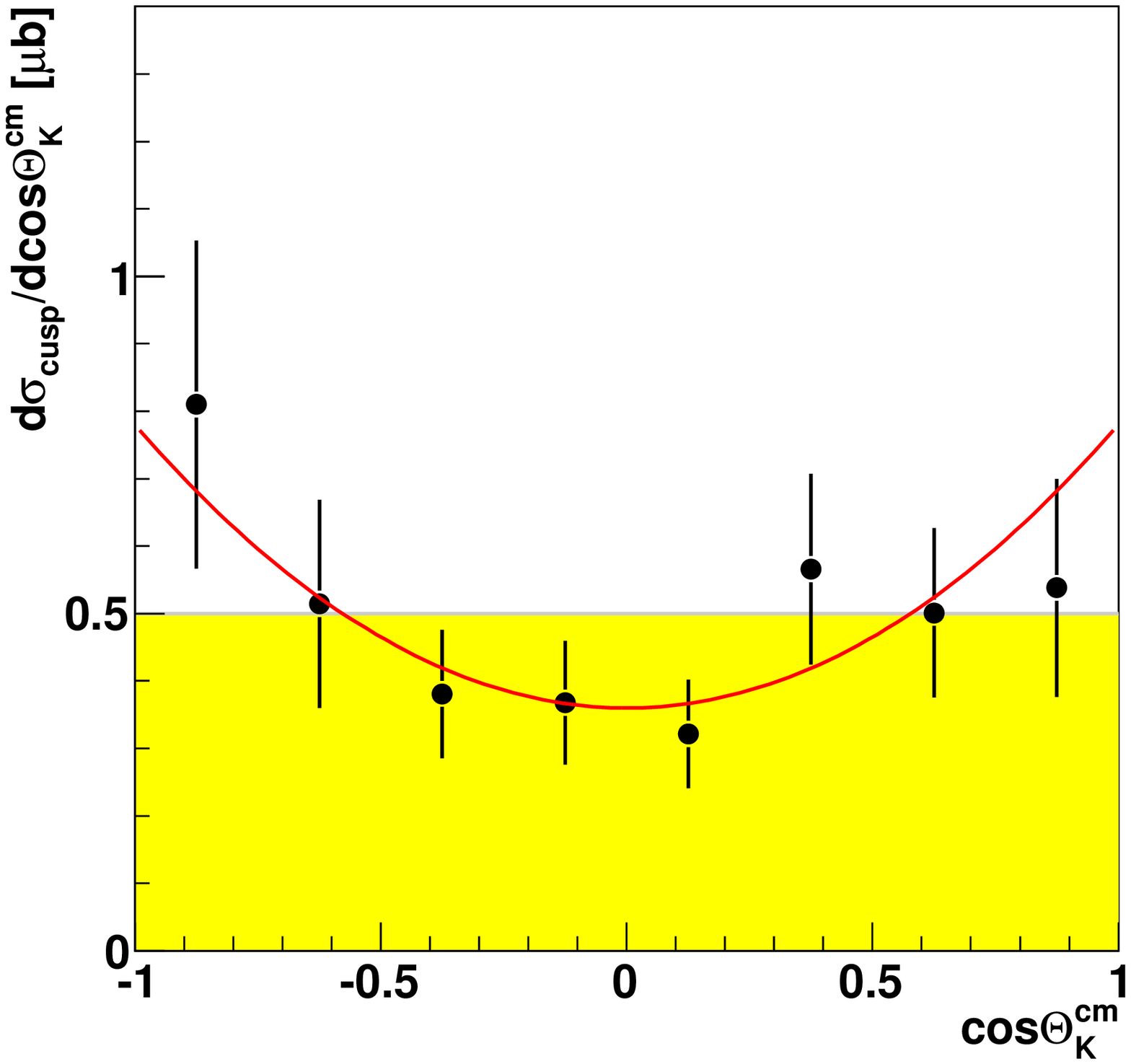}
\includegraphics[width=17pc]{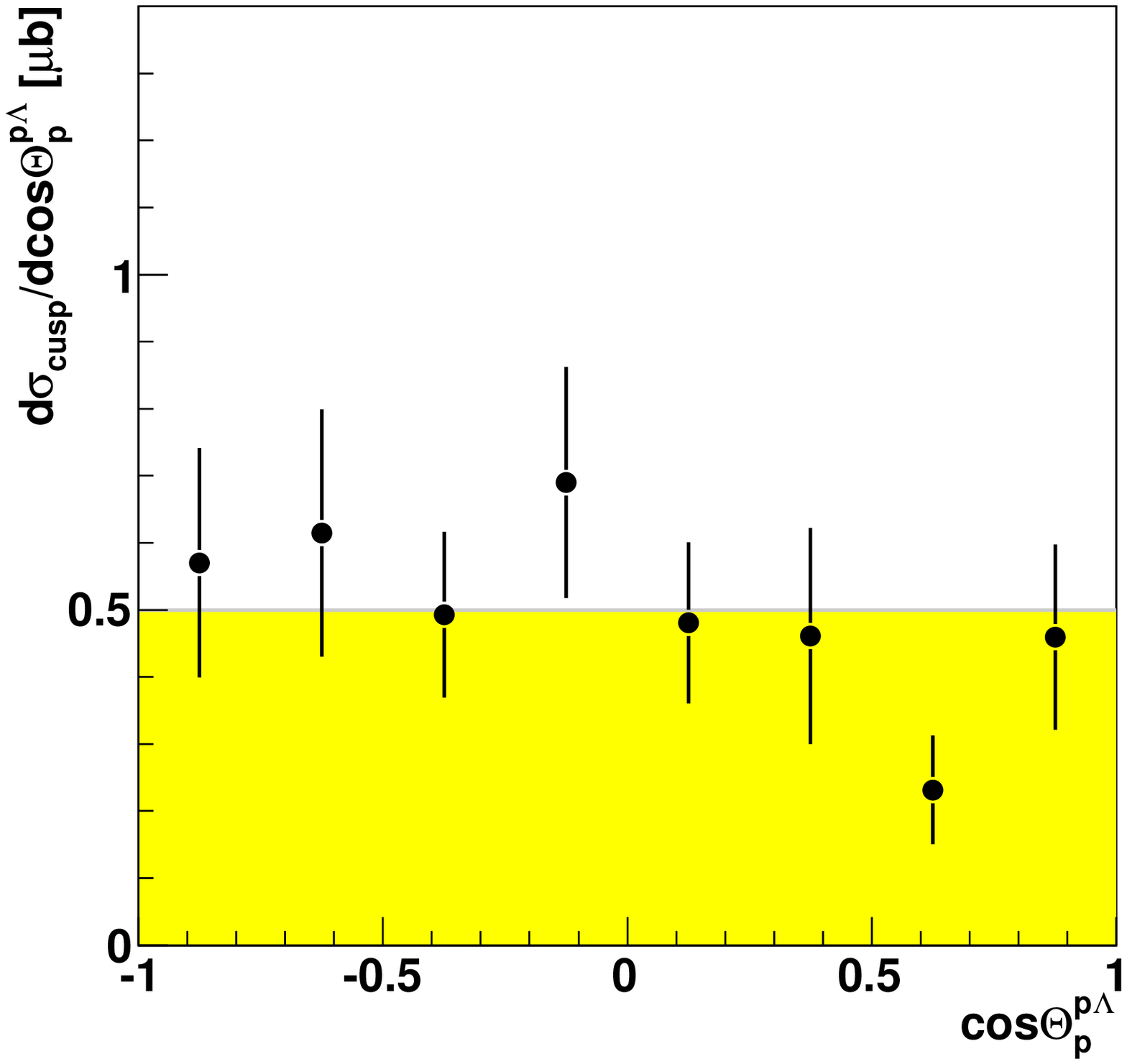} 
\end{center}
\caption{Distribution of the $\Sigma$N cusp over the $K^+$ angle in the
  center-of-mass system $\Theta_{K^+}^{cm} = 180^{\circ} - \Theta_{p\Lambda}^{cm}$
  ({\bf top}) and over the proton angle in the $p\Lambda$
  subsystem (Jackson frame) $\Theta_p^{p\Lambda}$ ({\bf bottom}). The plotted
  error bars include both statistical and systematic uncertainties.The shaded
  areas denote phase-space distributions. The solid curve in the top 
  figure represents a Legendre fit.
}
\end{figure}

The cusp scenario has a variety of consequences, which can be tested
experimentally by the cusp angular distributions.
 Such angular distributions are shown in Fig.~6.
 They have been obtained by subtraction of the background
as shown in Fig.~5, however, now for each angular
bin individually. 
{\it I.e.}, we split the double-differential cross section $d^2\sigma /
dM_{p\Lambda}dcos\Theta$ into eight equidistant angular bins of $dcos\Theta$,
fit then in each of these eight spectra a second-order polynomial to the data
in the regions outside the cusp and integrate the strength in the cusp region
above this polynomial line. This procedure of subtracting the background for
each angular bin individually has been applied successfully already in
previous TOF work, see {\it e.g.} Refs. \cite {DD,DD12}, for a detailed
presentation of this method see Ref. \cite{eta}.  
We note that a sideband background subtraction \cite{KE}
leads so similar results, however, it is in general not as reliable. 

Since the cusp effect is small compared to the
background originating from $N^*$ production, the cusp angular distributions
have now substantial uncertainties, both statistical and in particular
systematic. 
The systematic errors have been studied by varying the background description
with the second-order polynomial. As a result we find that the systematic
uncertainty is up to four times as big as the statistical uncertainty. The
latter is readily estimated from the fact that the cusp is just $1 / 20$ of
the total cross section. With a total of 30000 events and eight angular bins
we have on average only somewhat more than 100 events left in the cusp of an
angular bin spectrum, {\it i.e.} the statistical uncertainty is already in the
10$\%$ region.    
%

The condition of the $\Sigma$ being produced right at 
threshold means that the $\Sigma$ and $N$ are in relative $s$ wave with the
consequence that the spin-parity of the $\Sigma N$ system right
at threshold must be $J^P = 0^+$ or $1^+$. Hence, the $p\Lambda$ system
resulting from the $\Sigma N$ system can only be in relative $s$- or $d$-waves.
The observed proton angular distribution in the $p\Lambda$ subsystem (Jackson 
frame) within the $\Sigma N$ cusp is compatible with $s-$wave phase-space, 
see Fig. 6, bottom. A dominant $d$-wave contribution would lead to a 
strongly anisotropic distribution. We note that the observed distribution is
very different from the situation in the residual $pK^+\Lambda$ channel,
where the corresponding Jackson frame distribution exhibits strong $p$-wave
contributions -- see Fig. 7 in Ref.~\cite{DD}.

 The angular distribution of the $K^+$, shown in Fig. 6,  
top, is compatible with dominantly $s$-waves relative to the
$\Sigma N$ system, which is observed in our case as $p\Lambda$ system in the
$\Sigma N$ cusp. 
The $p$-wave component could be
obtained from the Legendre fit to the data according to the ansatz
\begin{center}
$d\sigma / d cos\Theta_K^{cm}~\sim~ a_0 + a_2(3~\cos^2\Theta_K^{cm} - 1)/2,~$

\end{center}
where the parameters 
$a_0=(0.50~\pm~0.10)$~$\mu$b and $a_2=(0.28~\pm~0.09)$~$\mu$b. 
 This result is different from the situation in the residual $pK^+\Lambda$
channel, however, very similar to that observed in the $pK^+\Sigma^0$ channel
at the same incident energy, see Fig. 10 and Table 5 in Ref.~\cite{DD},
and it is qualitatively similar to the $K^0$ cm angular distribution measured
in $pK^0\Sigma^+$ channel at 2.95 GeV/c~\cite{RG}. 

As mentioned in the introduction the only other data on the $\Sigma N$ cusp
in $pp$ induced $K$ production originate from inclusive single-arm 
magnetic spectrometer measurements at Saclay \cite{sieb} and COSY (HIRES
collaboration \cite{hires2}). In both cases essentially only a sharp increase
in the cross section with a slight indication of a bump is seen in
the $K^+$ missing mass spectrum at the $\Sigma p$ threshold. Since in these
inclusive measurements $\Sigma$ production cannot be separated from $\Lambda$
production, the trailing slope of the $\Sigma N$ cusp is not observed. 
If we compare these missing mass spectra to the $M_{p\Lambda}$
spectrum in Fig. 5 -- though the latter contains the integration over all
$K^+$ angles -- we see that at least qualitatively the shape of the
single-arm spectra is very close to that of the $M_{p\Lambda}$ spectrum up to
the maximum of the $\Sigma N$ cusp. This is true also for the height of the cusp
relative to the $pK^+\Lambda$ continuum left from the cusp. 

In Ref.~\cite{sieb} the cusp has been extracted from data 
at $T_p$~=~2.3~GeV - which is close to 
our energy - by subtracting phase space distributions for $\Lambda$ and
$\Sigma$ production -- see Fig. 5 in Ref.~\cite{sieb}. As a result of this
very crude treatment they obtain a bump at the cusp position, which indicates
a width of roughly 10 MeV/$c^2$ --  
%
at variance with our findings of a much
broader structure. 

%
%

Also, the width of 20 MeV/$c^2$ observed now in exclusive measurements is  
much larger than the value of 3 MeV/$c^2$ assumed in Ref.~\cite{hires2} for the
width of the $\Sigma N$ cusp.  Since the width of the $\Sigma N$ cusp affects
sensitively the value for the total cross section of the $pp\to nK^+\Sigma^+$ 
cross section extracted in Ref.~\cite{hires2} from the inclusive $K^+$
missing mass spectrum, the value of that cross section reduces substantially
when accounting for the much larger width of the cusp extreacted here.

The calculations of Laget, which account for both $\Lambda$ and $\Sigma$
production as well as the $\Sigma N$ cusp effect, are published only for
specific $K^+$ scattering angles. This leaves us merely with the possibility
of a 
very qualitative comparison. In Refs.~\cite{sieb,Laget} the calculations are
shown for a $K^+$ scattering angle of 10$^\circ$ at $T_p$ = 2.3~GeV. Since the
latter is close to the incident energy of this work, we may compare that
calculation directly to the $M_{p\Lambda}$ spectrum in Fig. 5. From
visual inspection we see that Laget's calculation qualitatively gives the
right order of magnitude for this effect, however, the calculated cusp effect
appears to be substantially narrower than observed in our data --- as already
mentioned above.

The cusp possibly may shed also new light onto the question about the $\Sigma
 N$-FSI. From differential and total cross section measurements of the $pp \to
 pK^+\Sigma^0$ reaction it has been concluded that there is no sizeable
$\Sigma^0 p$-FSI \cite{DD}.  The presence of the pronounced cusp observed
in $pp\to pK\Lambda$ needs a different interpretation, 
since such a structure points to a very strong
$\Sigma N \to \Lambda N$ transition. Therefore even at the threshold the
$\Sigma N$ interaction is strongly inelastic, which might well be the reason
of diminishing any strong final state distortion in the excitation function
for $pp\to N K\Sigma$. 
It is therefore particularly interesting to look into data on the
 $pp\to pK^0\Sigma^+$ reaction, where the $\Sigma N$ final state is purely 
isospin 3/2 and therefore does not couple to the $\Lambda N$ channel.
In fact, differential and total cross section data for this channel have been 
taken recently at TOF ~\cite{DD12}. 
They show no sign of any significant FSI effects.
Following the argumentation above this means that there is no sizeable FSI
in the I=3/2 channel -- different to the situation in the I=1/2 channel. A
thorough theoretical explanation of these findings is highly desirable.
For recent data on that channel see Ref.~\cite{DD12} and references therein.

\section{Summary}
\label{sec:5}
This work presents the first exclusive and kinematically complete
 measurements of the $\Sigma N$ cusp effect in the $pp \to
 pK^+\Lambda$ reaction and establishes this phenomenon in proton induced
 $\Lambda$ production for the first time.
The data exhibit a pronounced asymmetric shape of the cusp, which is gently
declining at its high-energy side. This is opposite to what is expected from
a Flatt\'{e} distribution. Due to the gentle fall-off at its high-energy side  
the cusp appears to be significantly broader than anticipated from theoretical
predictions. Whether this is indicative of a narrow resonance above the cusp
energy -- as speculated in earlier measurements of kaonic deuterium and as 
also possibly suggested by the high-resolution data presented here, can not be
decided at the present stage.

%
The measured angular distributions of the cusp point
 to $s$-waves between the kaon and the $p\Sigma$ system, 
 as well as dominantly $s$-waves between the subsequently emerging
 $\Lambda$ and proton. 
Detailed theoretical calculations
 for this cusp effect would be highly welcome in view of these new
 higher statistics measurements.

\section{Acknowledgments}

This work has been supported by BMBF, DFG
(Europ. Gra\-duiertenkolleg 683) and COSY-FFE (Forschungs\-zen\-trum
J\"ulich). We acknowledge valuable discussions with 
F.~Hinterberger, H.~Machner, A.~Sibirtsev, H.~Str\"oher and C.~Wilkin. 

\end{document}